\documentclass[%
 aip, 
 apl,%
 amsmath,amssymb,
 reprint,%
superscriptaddress, floatfix
]{revtex4-1}

\usepackage{graphicx}
\usepackage{dcolumn}
\usepackage{bm}

\begin{document}


\title{Demonstration of Surface Electron Rejection with\\
 Interleaved Germanium Detectors  for Dark Matter Searches}

\affiliation{Division of Physics, Mathematics, \& Astronomy, California Institute of Technology, Pasadena, CA 91125 USA} 
\affiliation{Fermi National Accelerator Laboratory, Batavia, IL 60510 USA}
\affiliation{Lawrence Berkeley National Laboratory, Berkeley, CA 94720 USA}
\affiliation{Department of Physics, Massachusetts Institute of Technology, Cambridge, MA 02139 USA}
\affiliation{Pacific Northwest National Laboratory, Richland, WA 99352 USA}
\affiliation{Department of Physics, Queen's University, Kingston, ON K7L 3N6, Canada}
\affiliation{Department of Physics, Santa Clara University, Santa Clara, CA 95053 USA}
\affiliation{SLAC National Accelerator Laboratory/Kavli Institute for Particle Astrophysics and Cosmology, 2575 Sand Hill Road, Menlo Park, CA 94025 USA}
\affiliation{Department of Physics, Southern Methodist University, Dallas, TX 75275 USA}
\affiliation{Department of Physics, Stanford University, Stanford, CA 94305 USA}
\affiliation{Department of Physics, Syracuse University, Syracuse, NY 13244 USA}
\affiliation{Department of Physics, Texas A \& M University, College Station, TX 77843 USA}
\affiliation{Departamento de F\'{\i}sica Te\'orica and Instituto de F\'{\i}sica Te\'orica UAM/CSIC, Universidad Aut\'onoma de Madrid, 28049 Madrid, Spain}
\affiliation{Department of Physics, University of California, Berkeley, CA 94720 USA}
\affiliation{Department of Physics, University of California, Santa Barbara, CA 93106 USA}
\affiliation{Department of Physics, University of Colorado Denver, Denver, CO 80217 USA}
\affiliation{Department of Physics, University of Evansville, Evansville, IN 47722 USA}
\affiliation{Department of Physics, University of Florida, Gainesville, FL 32611 USA}
\affiliation{Department of Physics, University of Illinois at Urbana-Champaign, Urbana, IL 61801 USA}
\affiliation{School of Physics \& Astronomy, University of Minnesota, Minneapolis, MN 55455 USA}

\author{R.~Agnese} \affiliation{Department of Physics, University of Florida, Gainesville, FL 32611 USA}
\author{A.J.~Anderson} \affiliation{Department of Physics, Massachusetts Institute of Technology, Cambridge, MA 02139 USA}
\author{D.~Balakishiyeva} \affiliation{Department of Physics, University of Florida, Gainesville, FL 32611 USA}
\author{R.~Basu~Thakur} \affiliation{Fermi National Accelerator Laboratory, Batavia, IL 60510 USA}\affiliation{Department of Physics, University of Illinois at Urbana-Champaign, Urbana, IL 61801 USA}
\author{D.A.~Bauer} \affiliation{Fermi National Accelerator Laboratory, Batavia, IL 60510 USA}
\author{A.~Borgland} \affiliation{SLAC National Accelerator Laboratory/Kavli Institute for Particle Astrophysics and Cosmology, 2575 Sand Hill Road, Menlo Park, CA 94025 USA}
\author{D.~Brandt} \affiliation{SLAC National Accelerator Laboratory/Kavli Institute for Particle Astrophysics and Cosmology, 2575 Sand Hill Road, Menlo Park, CA 94025 USA}
\author{P.L.~Brink} \affiliation{SLAC National Accelerator Laboratory/Kavli Institute for Particle Astrophysics and Cosmology, 2575 Sand Hill Road, Menlo Park, CA 94025 USA}
\author{R.~Bunker} \affiliation{Department of Physics, Syracuse University, Syracuse, NY 13244 USA}
\author{B.~Cabrera} \affiliation{Department of Physics, Stanford University, Stanford, CA 94305 USA}
\author{D.O.~Caldwell} \affiliation{Department of Physics, University of California, Santa Barbara, CA 93106 USA}
\author{D.G.~Cerdeno} \affiliation{Departamento de F\'{\i}sica Te\'orica and Instituto de F\'{\i}sica Te\'orica UAM/CSIC, Universidad Aut\'onoma de Madrid, 28049 Madrid, Spain} 
\author{H.~Chagani} \affiliation{School of Physics \& Astronomy, University of Minnesota, Minneapolis, MN 55455 USA}
\author{M.~Cherry} \affiliation{Department of Physics, Stanford University, Stanford, CA 94305 USA}
\author{J.~Cooley} \affiliation{Department of Physics, Southern Methodist University, Dallas, TX 75275 USA}
\author{B.~Cornell} \affiliation{Division of Physics, Mathematics, \& Astronomy, California Institute of Technology, Pasadena, CA 91125 USA}
\author{C.H.~Crewdson} \affiliation{Department of Physics, Queen's University, Kingston, ON K7L 3N6, Canada}
\author{P.~Cushman} \affiliation{School of Physics \& Astronomy, University of Minnesota, Minneapolis, MN 55455 USA}
\author{M.~Daal} \affiliation{Department of Physics, University of California, Berkeley, CA 94720 USA}
\author{P.C.F.~Di~Stefano} \affiliation{Department of Physics, Queen's University, Kingston, ON K7L 3N6, Canada}
\author{E.~Do~Couto~E~Silva} \affiliation{SLAC National Accelerator Laboratory/Kavli Institute for Particle Astrophysics and Cosmology, 2575 Sand Hill Road, Menlo Park, CA 94025 USA}
\author{T.~Doughty} \affiliation{Department of Physics, University of California, Berkeley, CA 94720 USA}
\author{L.~Esteban} \affiliation{Departamento de F\'{\i}sica Te\'orica and Instituto de F\'{\i}sica Te\'orica UAM/CSIC, Universidad Aut\'onoma de Madrid, 28049 Madrid, Spain} 
\author{S.~Fallows} \affiliation{School of Physics \& Astronomy, University of Minnesota, Minneapolis, MN 55455 USA}
\author{E.~Figueroa-Feliciano} \affiliation{Department of Physics, Massachusetts Institute of Technology, Cambridge, MA 02139 USA}
\author{J.~Fox} \affiliation{Department of Physics, Queen's University, Kingston, ON K7L 3N6, Canada}
\author{M.~Fritts} \affiliation{School of Physics \& Astronomy, University of Minnesota, Minneapolis, MN 55455 USA}
\author{G.L.~Godfrey} \affiliation{SLAC National Accelerator Laboratory/Kavli Institute for Particle Astrophysics and Cosmology, 2575 Sand Hill Road, Menlo Park, CA 94025 USA}
\author{S.R.~Golwala} \affiliation{Division of Physics, Mathematics, \& Astronomy, California Institute of Technology, Pasadena, CA 91125 USA}
\author{J.~Hall} \affiliation{Pacific Northwest National Laboratory, Richland, WA 99352 USA}
\author{H.R.~Harris} \affiliation{Department of Physics, Texas A \& M University, College Station, TX 77843 USA}
\author{J.~Hasi} \affiliation{SLAC National Accelerator Laboratory/Kavli Institute for Particle Astrophysics and Cosmology, 2575 Sand Hill Road, Menlo Park, CA 94025 USA}
\author{S.A.~Hertel} \affiliation{Department of Physics, Massachusetts Institute of Technology, Cambridge, MA 02139 USA}
\author{B.A.~Hines} \affiliation{Department of Physics, University of Colorado Denver, Denver, CO 80217 USA}
\author{T.~Hofer} \affiliation{School of Physics \& Astronomy, University of Minnesota, Minneapolis, MN 55455 USA}
\author{D.~Holmgren} \affiliation{Fermi National Accelerator Laboratory, Batavia, IL 60510 USA}
\author{L.~Hsu} \affiliation{Fermi National Accelerator Laboratory, Batavia, IL 60510 USA}
\author{M.E.~Huber} \affiliation{Department of Physics, University of Colorado Denver, Denver, CO 80217 USA}
\author{A.~Jastram} \affiliation{Department of Physics, Texas A \& M University, College Station, TX 77843 USA}
\author{O.~Kamaev} \affiliation{Department of Physics, Queen's University, Kingston, ON K7L 3N6, Canada}
\author{B.~Kara} \affiliation{Department of Physics, Southern Methodist University, Dallas, TX 75275 USA}
\author{M.H.~Kelsey} \affiliation{SLAC National Accelerator Laboratory/Kavli Institute for Particle Astrophysics and Cosmology, 2575 Sand Hill Road, Menlo Park, CA 94025 USA}
\author{S.A.~Kenany} \affiliation{Department of Physics, University of California, Berkeley, CA 94720 USA}\author{A.~Kennedy} \affiliation{School of Physics \& Astronomy, University of Minnesota, Minneapolis, MN 55455 USA}
\author{C.J.~Kenney} \affiliation{SLAC National Accelerator Laboratory/Kavli Institute for Particle Astrophysics and Cosmology, 2575 Sand Hill Road, Menlo Park, CA 94025 USA}
\author{M.~Kiveni} \affiliation{Department of Physics, Syracuse University, Syracuse, NY 13244 USA}
\author{K.~Koch} \affiliation{School of Physics \& Astronomy, University of Minnesota, Minneapolis, MN 55455 USA}
\author{B.~Loer} \affiliation{Fermi National Accelerator Laboratory, Batavia, IL 60510 USA}
\author{E.~Lopez~Asamar} \affiliation{Departamento de F\'{\i}sica Te\'orica and Instituto de F\'{\i}sica Te\'orica UAM/CSIC, Universidad Aut\'onoma de Madrid, 28049 Madrid, Spain} 
\author{R.~Mahapatra} \affiliation{Department of Physics, Texas A \& M University, College Station, TX 77843 USA}
\author{V.~Mandic} \affiliation{School of Physics \& Astronomy, University of Minnesota, Minneapolis, MN 55455 USA}
\author{C.~Martinez} \affiliation{Department of Physics, Queen's University, Kingston, ON K7L 3N6, Canada}
\author{K.A.~McCarthy} \affiliation{Department of Physics, Massachusetts Institute of Technology, Cambridge, MA 02139 USA}
\author{N.~Mirabolfathi} \affiliation{Department of Physics, University of California, Berkeley, CA 94720 USA}
\author{R.A.~Moffatt} \affiliation{Department of Physics, Stanford University, Stanford, CA 94305 USA}
\author{D.C.~Moore} \affiliation{Division of Physics, Mathematics, \& Astronomy, California Institute of Technology, Pasadena, CA 91125 USA}
\author{P.~Nadeau} \affiliation{Department of Physics, Queen's University, Kingston, ON K7L 3N6, Canada}
\author{R.H.~Nelson} \affiliation{Division of Physics, Mathematics, \& Astronomy, California Institute of Technology, Pasadena, CA 91125 USA}
\author{L.~Novak} \affiliation{Department of Physics, Stanford University, Stanford, CA 94305 USA}
\author{K.~Page} \affiliation{Department of Physics, Queen's University, Kingston, ON K7L 3N6, Canada}
\author{R.~Partridge} \affiliation{SLAC National Accelerator Laboratory/Kavli Institute for Particle Astrophysics and Cosmology, 2575 Sand Hill Road, Menlo Park, CA 94025 USA}
\author{M.~Pepin} \affiliation{School of Physics \& Astronomy, University of Minnesota, Minneapolis, MN 55455 USA}
\author{A.~Phipps} \affiliation{Department of Physics, University of California, Berkeley, CA 94720 USA}
\author{K.~Prasad} \affiliation{Department of Physics, Texas A \& M University, College Station, TX 77843 USA}
\author{M.~Pyle} \affiliation{Department of Physics, University of California, Berkeley, CA 94720 USA}
\author{H.~Qiu} \affiliation{Department of Physics, Southern Methodist University, Dallas, TX 75275 USA}
\author{R.~Radpour} \affiliation{School of Physics \& Astronomy, University of Minnesota, Minneapolis, MN 55455 USA}
\author{W.~Rau} \affiliation{Department of Physics, Queen's University, Kingston, ON K7L 3N6, Canada}
\author{P.~Redl} \affiliation{Department of Physics, Stanford University, Stanford, CA 94305 USA}
\author{A.~Reisetter} \affiliation{Department of Physics, University of Evansville, Evansville, IN 47722 USA}
\author{R.W.~Resch} \affiliation{SLAC National Accelerator Laboratory/Kavli Institute for Particle Astrophysics and Cosmology, 2575 Sand Hill Road, Menlo Park, CA 94025 USA}
\author{Y.~Ricci} \affiliation{Department of Physics, Queen's University, Kingston, ON K7L 3N6, Canada}
\author{T.~Saab} \affiliation{Department of Physics, University of Florida, Gainesville, FL 32611 USA}
\author{B.~Sadoulet} \affiliation{Department of Physics, University of California, Berkeley, CA 94720 USA}\affiliation{Lawrence Berkeley National Laboratory, Berkeley, CA 94720 USA}
\author{J.~Sander} \affiliation{Department of Physics, Texas A \& M University, College Station, TX 77843 USA}
\author{R.~Schmitt} \affiliation{Fermi National Accelerator Laboratory, Batavia, IL 60510 USA}
\author{K.~Schneck} \affiliation{SLAC National Accelerator Laboratory/Kavli Institute for Particle Astrophysics and Cosmology, 2575 Sand Hill Road, Menlo Park, CA 94025 USA}
\author{R.W.~Schnee} \affiliation{Department of Physics, Syracuse University, Syracuse, NY 13244 USA}
\author{S.~Scorza} \affiliation{Department of Physics, Southern Methodist University, Dallas, TX 75275 USA}
\author{D.~Seitz} \affiliation{Department of Physics, University of California, Berkeley, CA 94720 USA}\author{B.~Serfass} \affiliation{Department of Physics, University of California, Berkeley, CA 94720 USA}
\author{B.~Shank} \affiliation{Department of Physics, Stanford University, Stanford, CA 94305 USA}
\author{D.~Speller} \affiliation{Department of Physics, University of California, Berkeley, CA 94720 USA}
\author{A.~Tomada} \affiliation{SLAC National Accelerator Laboratory/Kavli Institute for Particle Astrophysics and Cosmology, 2575 Sand Hill Road, Menlo Park, CA 94025 USA}
\author{A.N.~Villano} \affiliation{School of Physics \& Astronomy, University of Minnesota, Minneapolis, MN 55455 USA}
\author{B.~Welliver} \affiliation{Department of Physics, University of Florida, Gainesville, FL 32611 USA}
\author{D.H.~Wright} \affiliation{SLAC National Accelerator Laboratory/Kavli Institute for Particle Astrophysics and Cosmology, 2575 Sand Hill Road, Menlo Park, CA 94025 USA}
\author{S.~Yellin} \affiliation{Department of Physics, Stanford University, Stanford, CA 94305 USA}
\author{J.J.~Yen} \affiliation{Department of Physics, Stanford University, Stanford, CA 94305 USA}
\author{B.A.~Young} \affiliation{Department of Physics, Santa Clara University, Santa Clara, CA 95053 USA}
\author{J.~Zhang} \affiliation{School of Physics \& Astronomy, University of Minnesota, Minneapolis, MN 55455 USA}

\collaboration{The SuperCDMS Collaboration} 
\noaffiliation


\begin{abstract}

The SuperCDMS experiment in the Soudan Underground Laboratory searches for dark matter with a 9-kg array of cryogenic germanium detectors. Symmetric sensors on opposite sides measure both charge and phonons from each particle interaction, providing excellent discrimination between electron and nuclear recoils, and between surface and interior events. Surface event rejection capabilities were tested with two $^{210}$Pb sources producing $\sim$130 beta decays/hr. In $\sim$800 live hours, no events leaked into the 8--115~ keV signal region, giving upper limit leakage fraction $1.7 \times 10^{-5}$ at 90\% C.L., corresponding to $< 0.6$ surface event background in the future 200-kg SuperCDMS SNOLAB experiment.

\end{abstract}


\maketitle


\begin{figure*}[hbtp]
\vspace{-5pt}
\begin{center}
\includegraphics[height=1.83in]{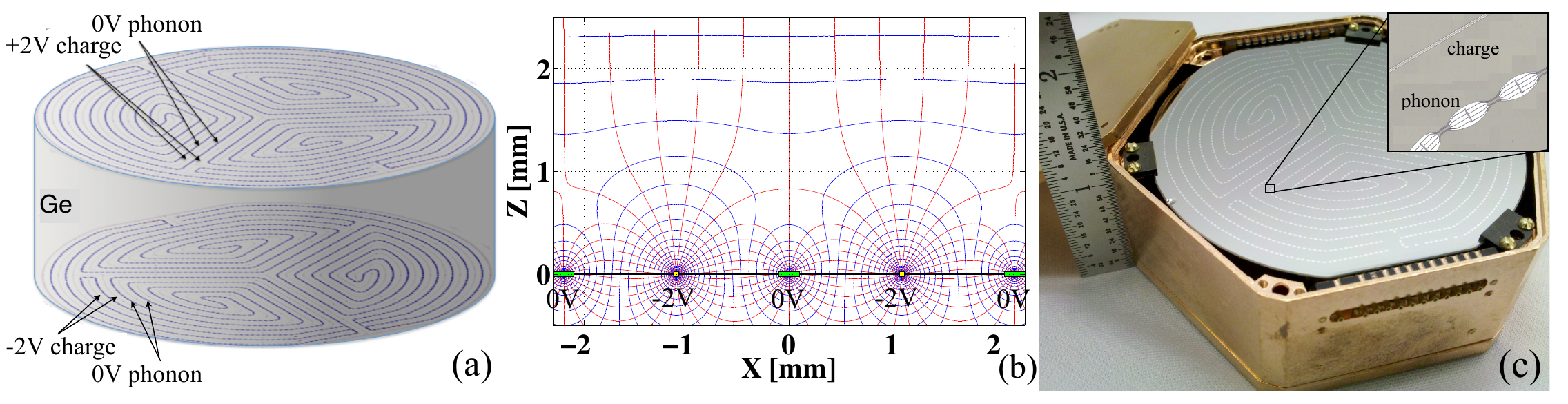}
\caption{\footnotesize (color online)
(a) Phonon and ionization sensor layout for iZIP detectors deployed at Soudan. The Ge crystal is 76~mm in diameter and 25~mm thick. Both faces are instrumented with ionization lines (one face with $+2$~V and the other with $-2$~V) that are interleaved with phonon sensors (0~V) on a $\sim$1~mm pitch. The phonon sensors are arranged to give 4 phonon readout channels for each face, an outer sensor surrounding three inner ones. 
(b) Magnified cross section view of electric field lines (red) and equipotential contours (blue) near the bottom face of a SuperCDMS iZIP detector. The $-2$~V ionization electrode lines (yellow) are narrower than the 0~V athermal phonon collection sensors (green).
(c) Fabricated iZIP detector in its housing.}
\label{fig:iZIP1}
\end{center}
\vspace{-15pt}
\end{figure*}

Weakly Interacting Massive Particles (WIMPs) are a generic class of candidates for the dark matter~\cite{wmap2008, lee1977, jungman} which is responsible for the formation of structure in our universe~\cite{blumenthal1984}. These Big Bang relic particles are particularly interesting because their possible existence is motivated by arguments both from cosmology and from particle physics.  Experiments are underway to detect WIMPs directly as they recoil off nuclei in terrestrial detectors~\cite{bertone2010}.

The approach of the Cryogenic Dark Matter Search (CDMS) is to maximize the information on each particle interaction using technology with excellent signal-to-noise and position information. These detectors with multiple readout channels have resulted in a series of robust experiments that have minimized unknown backgrounds. 
In addition to repeated improvements in sensitivity~\cite{r118prl,r118prd,r119prl,r119prd,prl102,c58science}, we have obtained constraints on annual modulation, inelastic dark matter interactions, axions, and electromagnetic interactions~\cite{Ann_Mod,cdmsiDM,axionprl,cdmsER}.

The CDMS technology senses both athermal phonons and ionization in Ge and Si crystals operated at $\sim$50~mK.
The low energy per excitation quantum in both ionization and phonons extends 
sensitivity to low-mass WIMPs~\cite{SUFLowthresh2010,SoudanLowthresh2011, pyle12}.
The nuclear recoils expected from WIMP interactions can be recognized through the measurement of the ionization yield, defined as the ratio of the measured ionization signal to the total recoil energy.
Separation between electron and nuclear recoils results in less than 1 electron recoil leaking into the nuclear band out of $1.7\times 10^6$ in the bulk volume of the detectors as measured by $^{133}$Ba calibration runs for recoil energies above 8~keVr,  where the `r' refers to the true recoil energy. Surface events taking place within a few tens of micrometers from the faces of the crystal, and events taking place in the outer radial portions of the detectors, can suffer from reduced ionization collection. These events thus have significantly degraded separation of electron and nuclear recoils. 

In order to reduce these dominant backgrounds for future experiments such as the 200 kg Ge SuperCDMS project planned for the SNOLAB laboratory, we have developed a new interleaved technology (iZIP)~\cite{brink06, pyle10}, which benefited from the EDELWEISS collaboration's experience~\cite{broniatowski2009}. These detectors have interleaved ionization and grounded phonon electrodes on both of the crystal faces, with a $+2$~V bias applied to the top ionization electrodes and $-2$~V applied to the bottom.
The ionization measurement is made by drifting the electron-hole pairs to electrodes on the crystal surface in a weak electric field ($\sim$0.5~V/cm).
The phonon measurement utilizes the advanced athermal phonon sensor technology developed for CDMS~II~\cite{Akerib_PRD}.
Athermal phonons propagating in the crystal interact with superconducting Al electrodes at the crystal surface, breaking Cooper pairs to form quasiparticles in the Al electrode.
Diffusion of quasiparticles to a tungsten ``Transition Edge Sensor'' (TES) increases the temperature and resistance of the TES, which is operated in the transition region between the superconducting and normal states.
The change in TES resistance under voltage bias is detected as a change in current using SQUID amplifiers.

\begin{figure}[htbp]
\begin{center}
\includegraphics[width=3.4in]{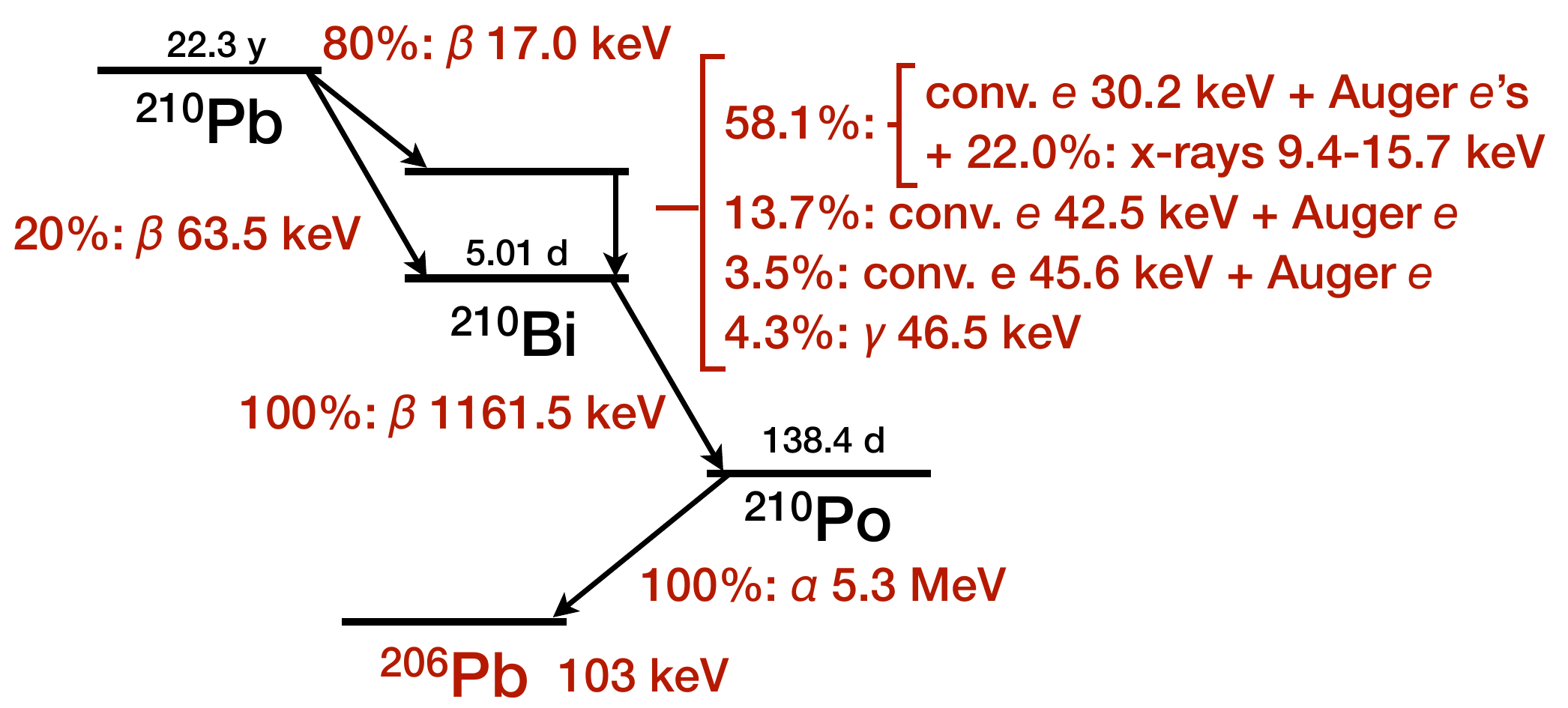}\\
\caption{\footnotesize Decay chain for $^{210}$Pb showing the most significant decays which end in a $^{206}$Pb nucleus from the $^{210}$Po alpha decay.
}
\label{fig:lead_decay}
\end{center}
\vspace{-5pt}
\end{figure}
  
\begin{figure*}[thbp]
\begin{center}
\includegraphics[height=2.5in]{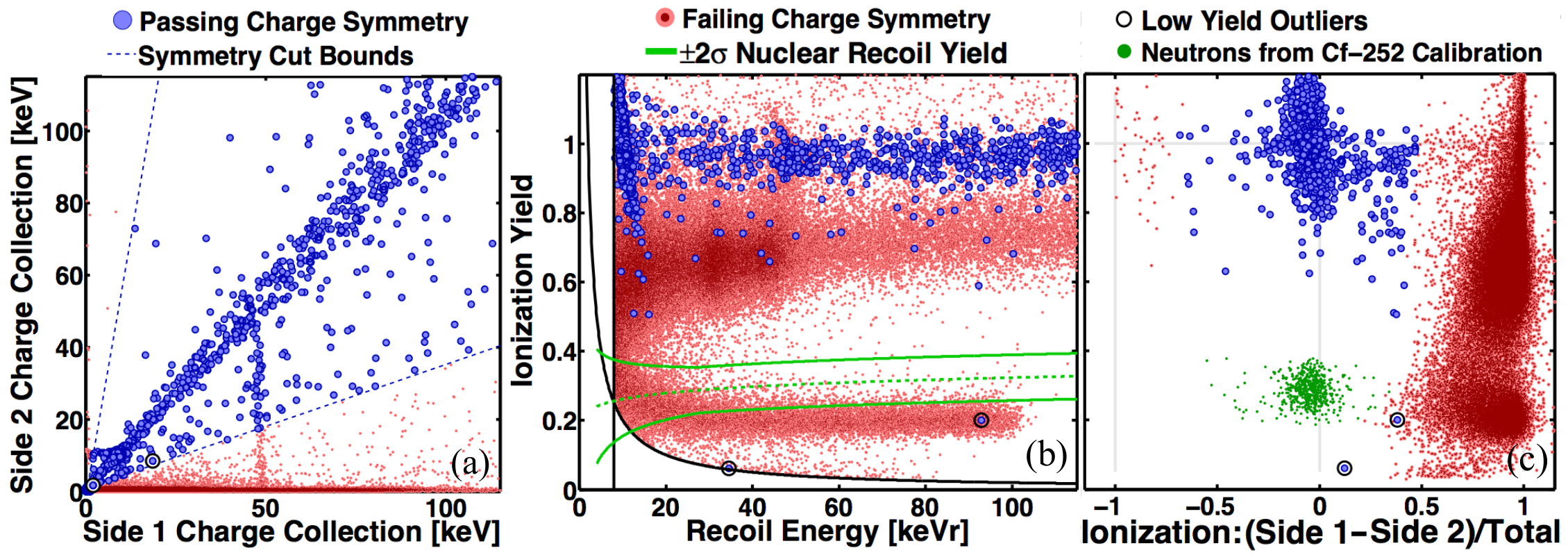}
\caption{\footnotesize  (color online)
All panels show the same data from $\sim$900 live hours of detector T3Z1 with the $^{210}$Pb source facing side 1. Clearly visible are the symmetric charge events (large blue dots) in the interior of the crystal, and the events that fail the symmetric charge cut (small red dots) including surface events from betas, gammas and lead nuclei incident on side 1 from the source. The two blue dots with circles around them are outliers that show a very low charge yield and just satisfy the symmetry requirement.
(a) The symmetry cuts (dotted blue lines) flare out near the origin so that events are accepted down to the noise wall. The band just below 50~keV is from the 46.5~keV gammas from the source.
(b) Ionization yield versus phonon recoil energy with $\pm 2 \sigma$ ionization yield range of neutrons indicated (area within green lines). The hyperbolic black line is the ionization threshold (2~keVee - `ee' for electron equivalent); the vertical black line is the recoil energy threshold (8~keVr). Electrons from $^{210}$Pb (below $\sim$60~keVr) and $^{210}$Bi (mostly above 60~keVr) are distinctly separated from $^{206}$Pb recoils (low yield, below $\sim$110~keVr).
(c) In addition to the data in (a) \& (b) this panel also shows nuclear recoils from neutrons from a $^{252}$Cf source (green, low yield). As bulk events these show a symmetric ionization response between side 1 and 2 like the bulk electron recoils at higher yield, and are thus nicely separated from charge-asymmetric surface events.}
\label{fig:soudan_demo}
\end{center}
\vspace{-20pt}
\end{figure*}

Figure~\ref{fig:iZIP1}a shows the electrode layout in use for SuperCDMS at the Soudan Underground Laboratory.
A detail of the resultant electric field near the surface of the Ge detector is shown in Fig.~\ref{fig:iZIP1}b, from which one may see that energy deposited deeper than $\sim$1~mm will liberate charges that drift to both faces of the crystal, whereas events near one surface will generate a charge signal read out only on that surface.
This asymmetry in charge collection significantly improves the ability of iZIP detectors to identify recoils that occur near the detector surface. Furthermore, the increased electric field near the surface improves charge collection for all surface events.

In addition to the interleaved electrode structure's rejection of near-surface events, the outermost ionization bias electrodes are instrumented as a veto guard ring.
An outer phonon channel enables estimation of event radial position, providing rejection of perimeter background events to lower recoil energies ($\sim$1~keV) than was possible in CDMS~II.
Such features of iZIP prototypes were studied extensively at the surface UC Berkeley (UCB) test facility~\cite{pyle10}.
The UCB studies yielded promising background rejection, but were limited by cosmogenic neutron background in the WIMP signal region.  In order to measure directly the background rejection for these events, 
$^{210}$Pb sources were installed in the Soudan Underground Laboratory experiment facing two detectors T3Z1(T3Z3), with the source facing the +2 V(-2 V) electrode. These sources were fabricated by the Stanford group~\cite{coleman09} using silicon wafers sealed in an aluminum box for 12 days with a 5~kBq $^{226}$Ra source producing $^{222}$Rn gas.  The silicon wafers were then etched with a standard wafer cleaning procedure and calibrated with an XIA ultra-low background alpha counter \cite{XIA}.  
The two deployed sources are nearly uniformly implanted with $^{210}$Pb to a depth of $\sim$58~nm and, by the decay chain~\cite{chiste2007,kibedi2008} shown in Fig.~\ref{fig:lead_decay}, give a total electron interaction rate of $\sim$130~events per hour in the 8--115~keVr region of interest.

As shown in Fig.~\ref{fig:soudan_demo}a, events taking place in the bulk of the detectors, such as the 10.4~keV Ge activation line, produce an ionization response that is symmetrically divided between the two faces of the iZIP. In contrast, surface betas from the source show a signal primarily on the side of the crystal facing the source.  
Events that take place in the outer radial regions of the detector, which can also suffer from reduced ionization yield,
were identified by comparing the ionization collected in the outer guard electrode to that collected in the inner electrode and
do not appear in the plot.

As seen in Fig.~\ref{fig:soudan_demo}b, surface betas from the $^{210}$Pb source populate a region of reduced ionization yield, which lies between the electron-recoil (ionization yield $\sim$1) and nuclear-recoil bands. The recoiling $^{206}$Pb nuclei from the $^{210}$Po alpha decay are also seen, with an ionization yield of $\sim$0.2 which is below the Ge nuclear recoil band because of reduced yield of Pb recoils in Ge versus Ge recoils in Ge. This low-yield band ends near the known 103 keV maximum recoil energy for the recoiling nucleus, thereby providing direct confirmation for our nuclear-recoil energy scale.
The iZIP's ability to reject surface events versus bulk nuclear recoils is demonstrated in Fig.~\ref{fig:soudan_demo}c.
 
In the energy band 8--115 keVr detector T3Z1(T3Z3) recorded 71,525 (38,178) electrons and
16,258 (7,007) $^{206}$Pb recoils in 905.5 (683.8) live hours at Soudan.  The expected background rates are $\sim$10,000 times lower and are neglected in this analysis.  A WIMP signal region is defined by the 2-sigma band around the mean yield measured for nuclear recoils (using a $^{252}$Cf neutron calibration source). A fiducial volume is defined based on ionization information, requiring that there is no charge signal above threshold in the outer ionization sensor and that the charge signal is symmetric with respect to the detector faces (blue points in Figs.~\ref{fig:soudan_demo}).  Using these criteria, no surface events are found leaking into the WIMP signal region above a recoil energy of 8~keVr. This fiducialization yields a spectrum-averaged acceptance  efficiency of $\sim$50\% in the energy range of 8--115~keVr for a $\sim$60 GeV/c$^2$ mass WIMP. The statistics-limited upper limit to the surface event leakage fraction is $1.7\times10^{-5}$ at 90\% C.L., similar to that found by EDELWEISS above a threshold of 15~keVr~\cite{broniatowski2009}.   For an exposure of 0.3 ton-yr with a 200 kg Ge SNOLAB experiment, this leakage fraction corresponds to an estimated leakage $<$ 0.6 events  at 90\% C.L. assuming the same $^{210}$Pb background contamination levels as achieved at Soudan.

\begin{figure}[htbp]
\begin{center}
\includegraphics[width=3.4in]{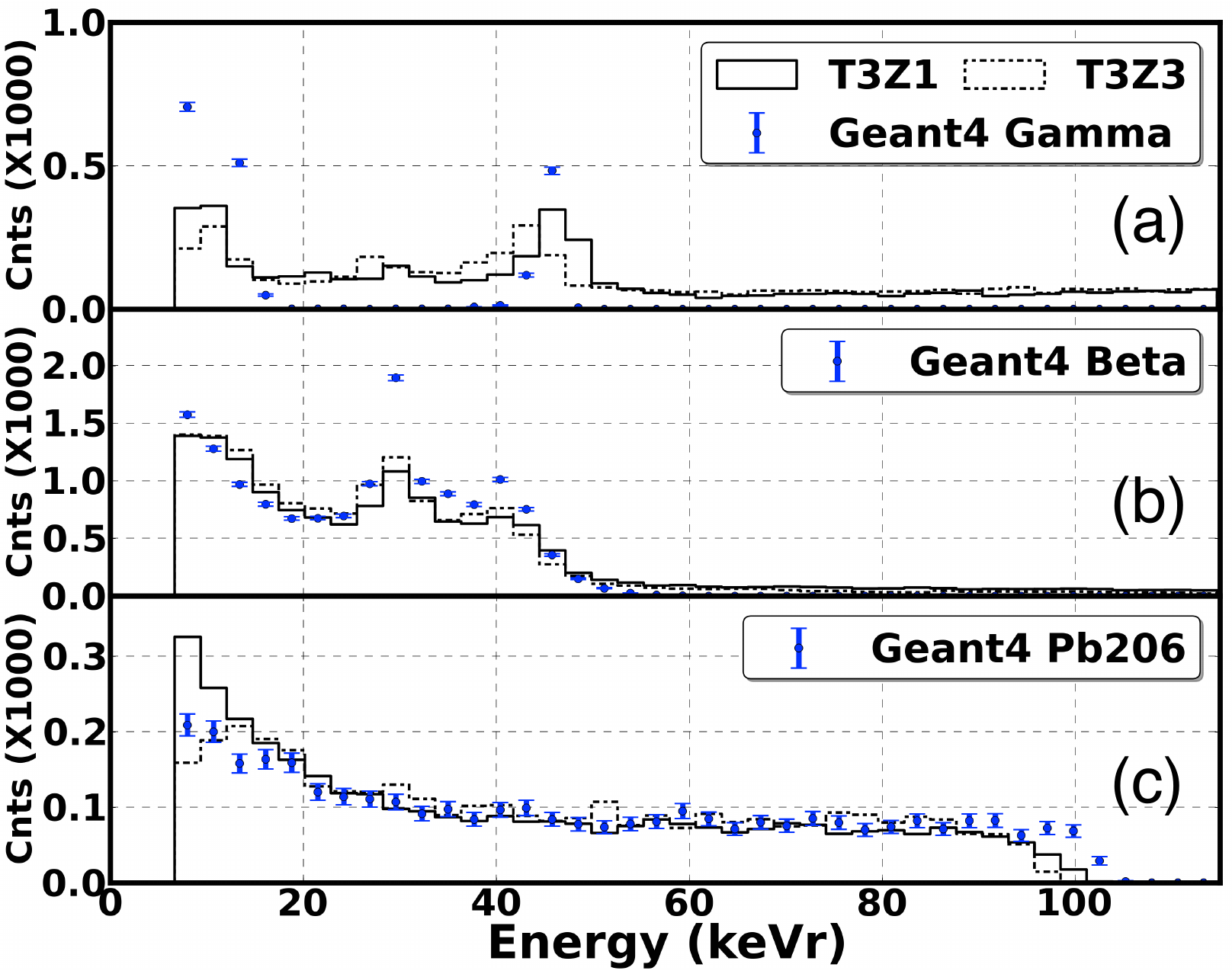}\\
\caption{\footnotesize 
Comparison of data from the two detectors with a Geant4 simulation. 
(a) Gamma electron recoil band (data with ionization yield $> 0.8$, including the not simulated continuous spectrum from bulk gamma interactions), compared with the simulated X-ray peaks at ~12 and 46.5 keV. (b) Beta band (data with yield 0.4-0.8), compared with the simulated beta decays from $^{210}$Pb or $^{210}$Po. (c) $^{206}$Pb recoil band (data with yield $< 0.4$), compared with the simulated $^{206}$Pb recoils from $^{210}$Po decay. 
Normalizations of the simulations in (a) and (b) were fixed by the normalization needed to match data and simulation in (c). Unlike the perfect simulation classifications, there is significant mixing in the data yield-based classification between ``gammas" and ``betas," and between ``betas" and ``recoils of $^{206}$Pb" at energies below $\sim$20~keV.}
\label{fig:lead_spectrum}
\end{center}
\vspace{-10pt}
\end{figure}

We analyzed the spectra from the two detectors, produced by the gammas, betas and lead recoils from the $^{210}$Pb sources. For this analysis, we used a Geant4 Monte Carlo~\cite{Agostinelli2003} adapted to improve the simulation of low energy ion implantation and other low energy processes~\cite{Mendenhall05}.  Fig.~\ref{fig:lead_spectrum} shows the comparison of the data against the simulation. The simulation modeled the intentional contamination with radon gas of the silicon source wafers, leading to surface adsorbed $^{214}$Po  which implants $^{210}$Pb, and then followed the subsequent decay chain in the source to $^{206}$Pb. There is generally good agreement between the Monte Carlo simulation and data, although Geant4's treatment of low energy X-rays, conversion electrons and Auger electrons requires more verification. Several known effects were not simulated, including: 1) surface roughness as in~\cite{kuzniak2012}, 2) the small fraction of ionized $^{218}$Po that plates out on the wafer prior to decay, potentially increasing the depth of the $^{210}$Pb implantation, 3) ionization yield to properly compare with data yield distributions, 4) beta events leaking into the lead recoil band, and 5) Frenkel pairs, lattice defects from nuclear recoils, which are predicted to cause small phonon energy quenching at $\sim$3\% level~\cite{lazanu2012}.  The upturn in the observed rate of $^{206}$Pb recoils at low energies may help explain some of the recoil events in the recent CRESST experiments~\cite{CRESSTII730kgd}.  
This upturn is more pronounced in T3Z1 than in T3Z3 because the lower charge threshold cut has higher efficiency below $\sim$15 keV.


As the analysis of the Soudan data is refined, we are exploring the use of phonon rise time and position reconstruction information to further improve rejection of surface events at low energies and reduce systematic uncertainties in the fiducial volume now defined using ionization measurements~\cite{hertel2012}.   
These phonon fiducial volume estimators can be used below the 8~keVr threshold used in this paper to augment the ionization-based estimate for the low-energy recoils expected from light WIMPs.

In conclusion, we have demonstrated that the new iZIP Ge detectors have sufficient surface electron rejection so that this background will be negligible for the current SuperCDMS Soudan experiment and contribute $< 0.6$ event background during a 0.3~ton-year exposure for the 200 kg SuperCDMS SNOLAB experiment.

\begin{acknowledgments}
The SuperCDMS collaboration gratefully acknowledges the technical assistance from Jim Beaty and the staff of the Soudan Underground Laboratory and the Minnesota Department of Natural Resources. These iZIP detectors are fabricated in the Stanford Nanofabrication Facility, which is a member of the National Nanofabrication Infrastructure Network sponsored by NSF under Grant ECS-0335765. This work is supported in part by the National Science Foundation (Grant Nos. AST-9978911, NSF-0847342, PHY-1102795, NSF-1151869, PHY-0542066, PHY-0503729, PHY-0503629, PHY-0503641, PHY-0504224, PHY-0705052, PHY-0801708, PHY-0801712, PHY-0802575, PHY-0847342, PHY-0855299, PHY-0855525, and PHY-1205898), by the Department of Energy (Con- tracts DE-AC03-76SF00098, DE-FG02-92ER40701, DE-FG02-94ER40823, DE-FG03-90ER40569, and DE-FG03-91ER40618, and DE-SC0004022), 
by NSERC Canada (Grants SAPIN 341314 and SAPPJ 386399), and by MULTIDARK CSD2009-00064 and FPA2012-34694. Fermilab is operated by Fermi Research Alliance, LLC under Contract No. De-AC02-07CH11359, while SLAC is operated under Contract No. DE-AC02-76SF00515 with the United States Department of Energy.

\end{acknowledgments}


\end{document}